\documentclass[titlepage]{article}
\usepackage{amssymb}
\usepackage{amsmath}
\usepackage{amsfonts}
\usepackage{graphicx}
\usepackage{epstopdf}

\newcommand{\chiqui}[1]{_{\mbox{\tiny{#1}}}}

\newcommand{\total}[2]{\frac{d#1}{d#2}}
\newcommand{\parcial}[2]{\frac{\partial#1}{\partial#2}}
\newcommand{\dtotal}[2]{\frac{d^{2}#1}{d#2^{2}}}

\newcommand{\metedos}{\varphi_{\Omega}}
\newcommand{\mb}{b_{\Omega}}
\newcommand{\ma}{a_{\Omega}}
\newcommand{\mkhh}{\chi_{\Omega}}

\newcommand{\under}{\underline}
\newcommand{\mitad}{\frac{1}{2}}
\newcommand{\e}{\epsilon}
\newcommand{\Om}{\Omega}
\newcommand{\om}{\omega}
\newcommand{\vphi}{\varphi}
\newcommand{\bx}{\under{x}}
\newcommand{\bn}{\under{n}}

\newcommand{\bra}[1]{\left\langle\, #1 \, \right|}
\newcommand{\ket}[1]{\left|\, #1\, \right\rangle}
\newcommand{\espe}[3]{\left\langle\, #1\, \left|\, #2\, \right|\, #3\, \right\rangle}

\newcommand{\expo}[1]{\exp\left\{#1\right\}}
\newcommand{\sign}{\text{sign}\,}

\newcommand{\lomas}{\ln_{+}z}

\title{Entanglement Entropy of Black Shells}
\author{J. Robel Arenas\thanks{e-mail: jrarenass@unal.edu.co.}\\ Observatorio Astron\'omico Nacional\\ Universidad Nacional de Colombia, Bogot\'a, Colombia \vspace{0.5cm} \\
J. Manuel Tejeiro\thanks{e-mail: jmtejeiros@unal.edu.co. Mailing Address: Observatorio Astron\'omico, Universidad Nacional de Colombia, Ciudad Universitaria- Calle 45, Cra.30, Bogot\'a, Colombia. Tel: 57 1 3165222. Fax: 57 1 3165383 }\\Observatorio Astron\'omico Nacional\\ Universidad Nacional de Colombia, Bogot\'a, Colombia}

\begin{document}

\maketitle

\begin{abstract}
We present a coherent account of how the entanglement interpretation, thermofield dynamical description and the brick wall formulations (with the ground state correctly identified) fit into a connected and self-consistent explanation of what Bekenstein-Hawking entropy is, and where it is located. \\

\end{abstract}

\section{Introduction}
The Bekenstein-Hawking entropy $S\chiqui{BH}$ has been derived in so many ways and interpreted from so many points of view. All existing derivations agree that the Bekenstein-Hawking entropy is proportional to surface area $A$,
\begin{equation}\label{1}
S\chiqui{BH}=\frac{A}{4(l_{pl})^2}\;. 
\end{equation}
 It is an entirely superficial result, in the literal sense that it refers solely to surface properties. So either, a profound holographic principle is at work (all information about the black hole interior is somehow imprints on the horizon)~\cite{ref:1} or existing derivations are ``superficial'', in the sense that they refer to an effective black shell entropy and don't probe the real black hole interior at all.  Following this last idea, in this paper we present a basic model of black shell from the existence of thermal energy strongly concentrated near the horizon (thermal energy ``wall'') with respect to an uniformly accelerated observer, according to the equivalence principle.
 
Among the different derivations of $S\chiqui{BH}$ that have been proposed to provide a microscopic explanation of the Bekenstein-Hawking Entropy~\cite{ANC2,EA2,EA3}, maybe the most promising and appropriate approach is that $S\chiqui{BH}$ is entanglement entropy, associated with observable and non-observable vacuum fluctuations correlated at the horizon with respect to an external observer. A program of this type was first clearly formulated by Bombelli et al.~\cite{BKLS1}. It was independently re-initiated by M. Srednicki~\cite{BKLS3} and by V. P. Frolov and I. D. Novikov~\cite{EA6}. From this approach the entanglement  entropy  proportional to the area of the dividing wall can be obtained. These striking results  of entanglement entropy are entirely general, by no means confined to systems in thermal equilibrium. Accordingly, they do not within themselves bear any clue as to the origins of the peculiarly thermal character of the Bekenstein-Hawking entropy. A comprehensive understanding of the Bekenstein-Hawking entropy requires a consistent blend of the entanglement interpretation with the thermofield-dynamical description~\cite{EU9,ETE1,ref:I}. Moreover, this allows one to say not only what this entropy is but where it is located. The resulting picture (which we elaborate in Sec.~3) is of a ``thermal atmosphere''  extending a few Planck lengths above the horizon.

On the other hand, the entanglement interpretation is implicit in, and is certainly closely related to the brick wall model introduced by 't Hooft ~\cite{BW1}, which will be shown below. In this sense, we use the modified brick wall model~\cite{BW2} and consider it as a model of black shell (a massive spherical shell compressed into a thin layer near its gravitational radius). This correct description (Sec.~3) shows that the thermal atmosphere is properly understood as excitations above an energetically depressed ground state (the Boulware state), and provides an accurate localization of these excitations in the Hartle-Hawking state. Because this localization involves \textit{differential} expectation values, the calculations are actually simplified, since the ultraviolet divergences cancel out.

It is thus our aim in the following sections to provide a coherent account of how the entanglement concept  thermofield dynamical description and the brick wall formulations fit into a connected and self-consistent  explanation of what Bekenstein-Hawking entropy is, in terms of a model of a massive spherical shell compressed into a thin layer near its gravitational radius, considering the Hartle-Hawking state as an effective Hartle-Hawking state for a modified brick wall model~\cite{BW2}.

\section{Thermally entangled Minkowski vacuum state}

Consider a real scalar field $\Phi$ defined on the geometrical background  with static metric given by 
\begin{equation}
\label{EU16-1}
ds^{2} = -f(r)\, d{\tau}^{2} + \frac{dr^{2}}{f(r)} + dz^{2} + dy^{2}\,,
\end{equation}
for $f(r)=r$.\\
One set of mode solutions of the corresponding wave equation for $R: x>|t|$ sector of Minkowski spacetime is the following:
\begin{equation}
\label{EU16-2}
\varphi_{\Omega}(\tau, x^{a}) = \varphi_{\Omega}(r)\,  e^{i(k_{\chiqui{1}} z +k_{\chiqui{2}} y)}\, \frac{1}{\sqrt{2|\omega| (2\pi)^2}}\, e^{-i\omega \tau}\, , 
\end{equation} 
with 
$\underline{x} \equiv x^{a} = (r, z, y)\, , 
\Omega \equiv\,\omega k_{\chiqui{1}}  k_{\chiqui{2}}$, where the modes $\varphi_{\Omega}(r)$ satisfies the equation
\begin{equation}\label{EU16-3}
r\total{}{r}\left(r\, \total{}{r}\varphi_{\Omega}(r)\right) +
\left[\omega^{2} - \left( k_{\chiqui{1}}^2+ k_{\chiqui{2}}^2 +m^{2}\right)f(r)
\right]\varphi_{\Omega}(r) = 0\, .
\end{equation}

Now  we introduce Killing-Boulware (KB)-modes $\metedos^{(\epsilon)}(x)$ for  $R$ and $L: x<|t|$ sectors
\begin{equation}\label{EU16-4}
\metedos^{(\epsilon)}(x) = \metedos(\underline{x}, \tau)\, \Theta_{\epsilon}(x),
\end{equation}
where $\Theta_{\epsilon}(x)$ is defined in terms of the null coordinates $u,v$,
\begin{equation} \label{ETE17}
\Theta_{\epsilon}(x) \equiv \frac{1}{2}\, \left\{\Theta(-\epsilon u) + \Theta(\epsilon
v)\right\}.
\end{equation}

According to thermofield dynamics, for a quantum description consider both quantization schemes,  Boulware-Rindler and Minkowski, that is
\begin{equation}
\Phi(x) = \sum_{\e, \Om}\ma^{(\e)}\, \mkhh^{(\e)}(x) = \sum_{\e, \Om}
\mb^{\e}\, \metedos^{(\e)}(x)\, ;\label{ETE59}
\end{equation}
\begin{equation}
\ma^{(\e)} = \mb^{(\e)}\, \cosh \chi - \mb^{(-\e)}\, \sinh \chi = e^{-iG}\,
\mb^{(\e)}\, e^{iG}\, , \label{ETE600}
\end{equation}
where,
\begin{equation}
G = G^{\dag} \equiv \mitad\, i\sum_{\e, \Om}\e(\om)\, \e\, \chi\,
\mb^{(\e)\dag}\, \mb^{(-\e)}  
\end{equation}
\begin{equation}
G= i\sum_{\substack{\Om \\ (\om > 0)}}\chi\, \left(\mb^{(+)\dag}\,
\mb^{(-)} - \mb^{(+)}\, \mb^{(-)\dag}\right) \equiv \sum_{\substack{\Om \\ (\om
> 0)}}G_{\Om}\, ; \label{ETE610} 
\end{equation}
\begin{equation}
\left[\ma^{(\e)},a_{\Om'}^{(\e')\dag}\right] = \left[\mb^{(\e)},
b_{\Om'}^{(\e')\dag}\right] = \e(\om)\, \e\, \delta_{\e \e'}\, \delta_{\Om
\Om'}\, ; \label{ETE62}
\end{equation}
with the vacuum states defined by
\begin{equation}
\ma^{(\e)
}\ket{0}_{\mbox{\tiny{M}}} = 0\, , \quad \mb^{(\e)}\ket{0}_{\mbox{\tiny{R}}} =
0\, , \quad (\om \e > 0)\, .\label{ETE63}
\end{equation}
According to \eqref{ETE600}, these vacuum states are formally linked by
\begin{equation}\label{ETE640}
\ket{0}_{\mbox{\tiny{M}}} = e^{-iG}\, \ket{0}_{\mbox{\tiny{R
}}}\, .
\end{equation}
From \eqref{ETE640} and \eqref{ETE610} we can write
\begin{equation}\label{EU16}
\ket{0}_{\mbox{\tiny{M}}} = Z^{-\mitad}\sum_{\bn}e^{-\mitad\, \beta\, E_{\bn}}\,
\ket{\bn}_{\mbox{\tiny{L}}}\, \ket{\bn}_{\mbox{\tiny{R}}}\,, 
\end{equation}
where, $\ket{\bn}_{\mbox{\tiny{BL}}}$ and $\ket{\bn}_{\mbox{\tiny{BR}}}$ are excitations of the parochial Boulware states $\ket{0}_{\mbox{\tiny{BL}}}$ and $\ket{0}_{\mbox{\tiny{BR}}},$ respectively, and
\begin{equation}
\bn \equiv \{n_{\Om}\, , \quad \text{for all $\Om$ with $\om > 0$}\}\, ; \notag \\
E_{\bn} = \sum_{\substack{\Om \\ \om > 0}}n_{\Om}\, \om\, ,\quad
Z=\sum_{\bn}e^{-\beta E_{\bn}}. 
\end{equation}
Also, from \eqref{EU16} we can define  Boulware state $\ket{0}\chiqui{BR}$  over the complete space as $\ket{0}\chiqui{M}$ depopulated of all Boulware modes  $\ket{\bn}\chiqui{BR}$ in $R$-sector,
\begin{equation}\label{EU17}
\ket{0}_{\mbox{\tiny{BR}}}\text{(over $L + R$)} = Z^{-\mitad}\sum_{\bn}e^{-\mitad\, \beta\, E_{\bn}}\, \ket{\bn}_{\mbox{\tiny{L}}}\, \ket{0}_{\mbox{\tiny{R}}}\, .
\end{equation}

\subsection{Hamiltonian Formulation}

The physical sense of the transformation \eqref{ETE600} is based  in the invariant action $S[\Phi]$ and invariant Hamiltonian $H$ under this transformation, according to the action 
\begin{equation}\label{ETE33}
S[\Phi] = 
\int_{-\infty}^{\infty}dt\, \left(\sum_{\epsilon}\epsilon\,
L^{(\epsilon)}(\Phi)\right)\, , 
\end{equation}
where the Lagrangian $L$ is given by
\begin{equation}\label{ETE34}
L = \sum_{\epsilon}\epsilon\, L^{(\epsilon)}(\Phi)\, ; \quad
L^{(\epsilon)}(\Phi) = \int \mathcal{L}^{(\epsilon)}(\Phi)\, d^{3}\under{x}\, ,
\end{equation}
\begin{equation}\label{ETE35}
\mathcal{L}^{(\epsilon)}(\Phi) = \frac{\sqrt{-g}}{2}\, \left\{-g^{00}\,
\Phi^{2},_{t} - \left(g^{ab}\, \Phi,_{a}\, \Phi,_{b} +
m^{2}\Phi^{2}\right)\right\}\Theta_{\epsilon}(x)\, , 
\end{equation}
with $g \equiv \det g_{\mu \nu}$.\\
Then, from \eqref{ETE34}, we can obtain the on-shell calculation for the expectation value of the local Hamiltonian operator  $H_{L}^{R}$ in the global quantum state $\ket{0}_{\mbox{\tiny{M}}}$
\begin{equation}\label{EU12}
\chiqui{M}\espe{0}{H_{L}^{R}}{0}{\chiqui{M}} = -Z^{-1}\sum_{\{n_{\om}\}}e^{-\beta E_{\bn}}\
E_{\bn} =  \langle\, E\, \rangle_{\beta},
\end{equation} 
where the local Hamiltonian  is given by
\begin{equation}\label{ETE84}
H_{L}^{R} = \int \mathcal{H}^{(+)}\, d^{3}\bx = \mitad
\sum_{\Om}|\om|\left(\mb^{(+)\dag}\, \mb^{(+)} + \mb^{(+)}\,
\mb^{^{+}\dag}\right)\ .
\end{equation} 

The resulting \eqref{EU12} means that a static observer in the region $R$ perceives  the vacuum state $\ket{0}\chiqui{M}$ as a thermally excited respecting vacuum state $\ket{0}\chiqui{R}$. We can interpret  it as the thermal feature of Minkowski vacuum for a local observer, due to his restriction to region $R$ and the presence of an event horizon; whose effects are a shift of his ground state bellow the global Minkowski ground state and it perceived as thermally excited, respectively. That is, Minkowski vacuum is perceived by this restricted observer as excited thermally above his ground state.

\subsection{Minkowski thermal energy wall}

 In order to show the existence of a thermal energy ``wall'' associated to the properties of the physical vacuum in the vicinity of the horizons, let us calculate the expectation value for the component $T_{00}(x, x')$ of the stress-energy tensor with respect to Minkowski and Rindler vacuum states for a scalar field. In general, for a scalar field, the expectation value of the stress-energy tensor $T_{\alpha \beta}(x)$ is given by
\begin{equation}\label{20}
\langle T_{\alpha \beta}(x, x')\rangle\,  = \mathcal{D}_{\alpha \beta'}\,
W(x, x'), \\
\end{equation}
where $W(x, x')$ is the Wightman function and
\begin{equation}\label{21}
\mathcal{D}_{\alpha \beta'} = \partial(_\alpha\, \partial_{\beta'}) -
\mitad\, g_{\alpha \beta}\left(\partial^{\gamma}\partial_{\gamma'} +
m^{2}\right).\, 
\end{equation}
Then,
\begin{equation}\label{22}
\lim_{x \to x'} \langle\, T_{\chiqui{0}}^{\chiqui{0}}(x, x')\, \rangle\chiqui{M-R} =
\partial^{\chiqui{0}}\, \partial_{\chiqui{0'}}\, W\chiqui{M-R}.
\end{equation} 
\begin{equation}\label{EU16-5}
\left.\partial_{0}\, \partial_{0'}(W\chiqui{M} - W\chiqui{R})(x, x')\right|_{x'
= x} \equiv \left. \partial_{0}\, \partial_{0'}\, W\chiqui{M-R}(x,
x')\right|_{x' = x},
\end{equation}
where $W\chiqui{M-R}=W\chiqui{M} - W\chiqui{R}$ is the difference between Wightman functions for Minkowski and Rindler vacuum states, respectively.
\begin{align}
\left. \partial_{0}\, \partial_{0'}\, W\chiqui{M-R}(x, x')\right|_{x' = x}
&= \sum_{\Om}\frac{1}{e^{\beta |\om|} - 1}\, \partial_{0}\, \partial_{0'}\,
\left. \metedos^{(+)}(x)\, \metedos^{(+)*}(x')\right|_{x' = x} \notag \\
&= \sum_{\Om}\frac{\frac{1}{8\pi^2}\, |\om|}{e^{\beta |\om|} - 1}\,
\left|\metedos(\bx)\right|^{2}\, ,\label{EU16-6}
\end{align}
where the functions $\metedos(\bx)$ are
\begin{equation}\label{EU16-7}
\varphi_{\Omega}(\bx) = \frac{1}{2\pi}\, \varphi_{\Omega}(r)\,  e^{i(k_{\chiqui{1}} z +k_{\chiqui{2}} y)}\,.
\end{equation}
then, 
\begin{equation}\label{EU16-8}
\left. \partial_{0}\, \partial_{0'}\, W\chiqui{M-R}(x, x')\right|_{x' = x} = \sum_{\Om}\frac{\frac{1}{8\pi^2}\, |\om|}{e^{\beta |\om|} - 1}\,
\left|\metedos(r)\right|^{2}\,,
\end{equation}
\begin{equation}\label{EU16-8}
\left. \partial_{0}\, \partial_{0'}\, W\chiqui{M-R}(x, x')\right|_{x' = x} = \int_{0}^{\infty}d\om\,\frac{\frac{1}{4\pi^2}\, \om}{e^{\beta \om} - 1}\,\sum_{k_{\chiqui{1}}}\,\sum_{k_{\chiqui{2}}}
\left|\metedos(r)\right|^{2}\,. 
\end{equation}
Under WKB approximation, with
\begin{equation}\label{EU16-9}
\varphi(r) = f^{-1}\, \psi(r)\, ,
\end{equation}
the equation \eqref{EU16-3} leads to the differential equation
\begin{align}
&\left[\dtotal{}{r} + k^{2}(r; k_{\chiqui{1}},k_{\chiqui{2}})\right]\psi_{wk_{\chiqui{1}}k_{\chiqui{2}}}(r) = 0\,
,\label{EU16-10}
\intertext{with}
&k^{2} \equiv \frac{1}{f}\left\{\frac{\om^{2}}{f} - (k_{\chiqui{1}}^2+ k_{\chiqui{2}}^2) -m^{2} + f^{-1}\right\}\, .\label{EU16-11}
\end{align}
Then, we obtain
\begin{align}
&\partial_{0}\, \partial_{0'}\, W\chiqui{M-R} =
\frac{1}{2\pi^{2}}\int_{0}^{\infty}\frac{\om^{2}\, d\om}{e^{\beta \om} -
1}\, \int_{0}^{p\chiqui{\text{max}}} \frac{p\,dp}{f|k|}\, ,\label{EU16-12}
\intertext{where we approximate, for large $k_{\chiqui{1}},k_{\chiqui{2}}$,} \notag
&\qquad \qquad \sum_{k_{\chiqui{1}}}\,\sum_{k_{\chiqui{2}}}() \approx \int ()\,dp\, ,\quad \text{also}\quad
\sum_{\eta = \pm}() = 2\,().
\intertext{Moreover, $k\chiqui{\text{max}}(\om, r)$ was defined according to} 
&\qquad \qquad \qquad \qquad k^{2}(r; \om, k\chiqui{\text{max}}) = 0\, ,
\intertext{that is ,} 
&\qquad k_{\chiqui{1\,max}}^2+ k_{\chiqui{2\,max}}^2 =
\frac{\om^{2}}{f} - m^{2} + f^{-1}\equiv p^{2}\,
.\label{EU16-13}\\
&\qquad \int_{0}^{p\chiqui{\text{max}}} \frac{p\,dp}{f|k|} = \frac{\sqrt{f}\,p}{f}\,
.\label{EU16-14}
\end{align}
Thus, we finally obtain
\begin{equation}
\partial^{0}\, \partial_{0'}\, W\chiqui{M-R} =
\int_{0}^{\infty}\frac{E}{e^{\frac{E}{T}} - 1}\, \frac{4\pi p^{2}\,
dp}{(2\pi)^{3}}\, , \label{EU16-15}
\end{equation}
where the extra term in $-T_{0}^{0}$ can be shown to be zero to the same approximation.\\
This is the expected thermal expression for energy density of a hot scalar field, strongly concentrated near the horizon under WKB approximation, which is good near the horizon.

\section{Entanglement entropy model of black shells}

From the existence of thermal energy strongly concentrated near the horizon (thermal energy ``wall'') with respect to a uniformly accelerated observer, according to the equivalence principle, we can expect that identical thermal effects will occur, not only near the horizon of a very massive black hole but  also near the exterior of a starlike object with a reflecting surface, compressed to nearly (but not quite) its gravitational radius. In particular, the thermal description, summarized by  expression~(\ref{EU16-15}), is the same as the one established for the modified brick wall model~\cite{BW2}. Just as the Minkowski vacuum, according to the description above, is explainable to a uniformly accelerated observer as a thermal excitation above his negative-energy (Rindler) ground state, close to brick wall a delicate cancellation between a large thermal energy and an equally large and negative ground state energy is manifested. Following this model we may approximate the total stress-energy (ground state + thermal excitations) ${(T_{\alpha \beta})}\chiqui{H}$,  near the wall, to the Hartle-Hawking stress-energy ${(T_{\alpha \beta})}\chiqui{HH}$:
\begin{equation}
{(T_{\alpha \beta})}\chiqui{HH} \approx {(T_{\alpha \beta})}\chiqui{H} = {(T_{\alpha \beta})}\chiqui{B} + {(\Delta T_{\alpha \beta})}\chiqui{therm}
\end{equation}	
i.e., effectively the Hartle-Hawking stress-energy, which is bounded and small for large masses. So, in very good approximation for this case, near the wall we can use  Hartle-Hawking state as an effective Hartle-Hawking state.\\
In this context  we use Hartle-Hawking state below, for the model of a massive spherical shell compressed into a thin layer near its gravitational radius.

\subsection{Thermofield dynamics of black shells}

We consider the generic situation of a real scalar field $\Phi$ propagating on the geometrical background with static metric given by
\begin{equation}\label{ETE1}
ds^{2} = g_{\chiqui{00}}\, dt^{2}+ g_{ab}\, dx^{a}\, dx^{b}\, .
\end{equation}
Such a field $\Phi$ satisfies the field equation  
\begin{equation}\label{ETE2*}
( \Box - m^{\chiqui{2}})\, \Phi = 0\, , 
\end{equation}
where  d'Alembertian operator $\Box$ is expressed by
\begin{equation}\label{ETE3}
\Box = (-g)^{\chiqui{-1/2}}\partial_{\chiqui{0}}\left[(-g)^{\chiqui{1/2}} g^{\chiqui{00}} \partial_{\chiqui{0}}\right]
+ (-g)^{\chiqui{-1/2}}\partial_{a}\left[(-g)^{\chiqui{1/2}} g^{ab}\partial_{b}\right], 
\end{equation}
with $g \equiv \det g_{\mu \nu}$.

In particular, for $R$ sector in the Schwarzschild spacetime maximally extended ($R: Z>|T|$)   with static metric
\begin{equation}\label{ETE1A}
ds^{2} = -f(r)\, dt^{2} + \frac{dr^{2}}{f(r)} + r^{2}\, d\Om^{2},
\end{equation}
one set of mode solutions of Eq.~(\ref{ETE2*}) is the following:
\begin{equation}\label{ETE4}
\varphi_{\Omega}(t, x^{a}) = \varphi_{\Omega}(r)\, Y_{lm}(\theta,
\varphi)\, \frac{1}{\sqrt{2|\omega|}}\, e^{-i\omega t},
\end{equation}
where
\begin{equation}\notag
\underline{x} \equiv x^{a} = (r, \theta, \varphi)\, , \Omega \equiv
\omega l m.
\end{equation}
Substituting Eq.~(\ref{ETE4}) into Eq.~(\ref{ETE2*}) it is obtained
\begin{equation}\label{ETE5}
\frac{1}{r^{2}} \total{}{r}\left(r^{2} f(r)
\total{\varphi_{\Omega}(r)}{r}\right) + \left(\frac{\omega^{2}}{f(r)} -
\frac{l(l + 1)}{r^{2}} - m^{2}\right)\varphi_{\Omega}(r) = 0.
\end{equation}

There are two special solutions $\varphi_{\Omega \pm}(r)$ for the equation \eqref{ETE5}, defined by the boundary condition 
\begin{align}
&\varphi_{\Omega \pm}(r) \approx e^{\pm i\omega r_{*}}\, , \label{ETE8}\\
\text{for}\quad &r \to r_{0}\quad \text{(event horizon)} \notag \\
\text{or}\quad &r_{*} \to -\infty\quad \text{with $r_{*}$ defined by}  \notag \\
&dr_{*} \equiv\frac{dr}{f(r)},
\end{align}
which yield modes that are outgoing and ingoing modes on the horizon in $R$ sector. So, $\Omega \equiv
\omega k l m$ for $k \equiv \pm 1.$

Now, in order to concisely write modes for $R$ and $L$ sectors ($R: Z>|T|$ and $L: Z<|T|$), consider t-modes $\metedos(\underline{x}, t)$ for $R$ sector and Killing-Boulware (KB)-modes $\metedos^{(\epsilon)}(x)$ for  $R$ and $L$ sectors:
\begin{align}
\metedos(\underline{x}, t) &= \metedos(\underline{x})\,
\frac{1}{\sqrt{2|\omega|}}\, e^{-i\omega t}\, , \label{ETE15}\\
\metedos^{(\epsilon)}(x) &= \metedos(\underline{x}, t)\,
\Theta_{\epsilon}(x)\, ,\label{ETE16}\\
\intertext{where}
\Theta_{\epsilon}(x) &\equiv \frac{1}{2}\, \left\{\Theta(-\epsilon U) + \Theta(\epsilon
V)\right\}\, , \label{ETE17}
\intertext{ $\Theta$ is the unit step function,  $t$ runs backwards in $L$ sector ($\epsilon = -1$) and $U,V$, are the Kruskal times (see Appendix A).} \notag
\end{align}
With the K-G products
\begin{align}
\left(\metedos, \varphi_{\Omega'}\right) &= \epsilon(\omega)\,
\delta_{\Omega \Omega'},\label{ETE18}\\
\left(\metedos^{(\epsilon)}, \varphi_{\Omega'}^{(\epsilon')}\right) &=
\epsilon\, \epsilon(\omega)\, \delta_{\Omega \Omega'}\, \delta_{\epsilon
\epsilon'}\, , \label{ETE19}\\
\intertext{and}
\epsilon(\omega) &\equiv \text{sign}(\omega), \notag \\
\delta_{\Omega \Omega'} &= \delta(\omega - \omega')\, \delta_{k k'}\,
\delta_{l l'}\, \delta_{m m'}\, ,\notag \\
k &\equiv \pm 1. \notag
\end{align}
These (KB)-modes share the same static metric \eqref{ETE1A} but with different sign for the time coordinate. More precisely, (KB)-modes are positive frequency modes in Killing time Epsilon t in each of the sectors $L$, $R$.

For a complete set of modes given by Eq.~(\ref{ETE16}), we have the completeness relation
\begin{equation}\label{ETE27}
\frac{1}{2}\, \gamma(\under{x})\, \sum_{\Omega}\metedos(\under{x})\,
\metedos^{*}(\under{x}') = \delta^{3}(\under{x} - \under{x}')\, , 
\end{equation}
and the orthogonality relation
\begin{equation}\label{ETE28}
 \int d^{3}\bx\, \gamma(\bx)\, \varphi_{\Om'}^{*}(\bx)\,
\metedos(\bx) = \delta_{\Om \Om'} + \delta_{\bar{\Om} \Om'},
\end{equation}
where,
\begin{equation}\notag
 \bar{\Om} = -\om,lmk \nonumber.
\end{equation}

On the other hand, consider Kruskal-Hartle-Hawking (KH$^{2}$)-modes $\mkhh^{(\epsilon)}(x)$ which are positive frequency modes in Kruskal time  $U,V$ over complete manifold; i.e.  $\mkhh^{(\epsilon)}(x)$ is positive frequency in kruskal time if $\omega\, \epsilon > 0$, and $\mkhh^{(\epsilon)}(x)$ is negative frequency in Kruskal time if $\omega\, \epsilon < 0$ (See Appendix A).
These modes satisfy the orthogonality relation
\begin{equation}\label{ETE29}
\left(\mkhh^{(\epsilon)}, \chi_{\Omega'}^{(\epsilon')}\right) = \epsilon\,
\epsilon(\omega)\, \delta_{\Omega \Omega'}\, \delta_{\epsilon \epsilon'},
\end{equation}
and are connected with the modes  $\metedos^{(\epsilon)}(x)$ by the Bogolubov relation
\begin{equation}\label{ETE30}
\mkhh^{(\epsilon)}(x) = \metedos^{(\epsilon)}(x)\, \cosh \chi + \metedos^
{(-\epsilon)}(x)\, \sinh \chi,
\end{equation}
 where $\chi$ is defined in terms of
\begin{equation}
 \tanh \chi = e^{-\pi |\omega|/\kappa_{\chiqui{0}}}, \label{ETE31}
 \end{equation}
 and  $\kappa_{\chiqui{0}}$ is the surface gravity. 
 From equations  \eqref{ETE16},  \eqref{ETE30} and  \eqref{eq:MFP15} it is also calculated 
\begin{equation}\label{ETE32}
\mkhh^{(\epsilon)}(x) = \sqrt{\frac{\sinh \chi \, \cosh \chi}{2|\omega|}}\,
\metedos(\under{x})\, e^{-i\omega t_{\epsilon(\omega)\, \epsilon}}.
\end{equation}

For a quantum description consider both quantization schemes, for Boulware and Hawking-Hartle, that is, for the modes \eqref{ETE16} and \eqref{ETE32}, respectively,
\begin{equation}
\Phi(x) = \sum_{\e, \Om}\ma^{(\e)}\, \mkhh^{(\e)}(x) = \sum_{\e, \Om}
\mb^{(\e)}\, \metedos^{(\e)}(x)\,,\label{ETE59-1}
\end{equation}
where, from Eq.~(\ref{ETE30})
\begin{equation}
\ma^{(\e)} = \mb^{(\e)}\, \cosh \chi - \mb^{(-\e)}\, \sinh \chi = e^{-iG}\,
\mb^{(\e)}\, e^{iG}\, , \label{ETE60}
\end{equation}
and
\begin{align}
G &= G^{\dag} \equiv \mitad\, i\sum_{\e, \Om}\e(\om)\, \e\, \chi\,
\mb^{(\e)\dag}\, \mb^{(-\e)} \notag \\
&= i\sum_{\substack{\Om \\ (\om > 0)}}\chi\, \left(\mb^{(+)\dag}\,
\mb^{(-)} - \mb^{(+)}\, \mb^{(-)\dag}\right) \equiv \sum_{\substack{\Om \\ (\om
> 0)}}G_{\Om}.\label{ETE61} 
\end{align}
Now, we postulate commutation relations
\begin{equation}
\left[\ma^{(\e)},a_{\Om'}^{(\e')\dag}\right] = \left[\mb^{(\e)},
b_{\Om'}^{(\e')\dag}\right] = \e(\om)\, \e\, \delta_{\e \e'}\, \delta_{\Om
\Om'}\, ; \label{ETE62}
\end{equation}
with the vacuum states defined by
\begin{equation}
\ma^{(\e)
}\ket{0}_{\mbox{\tiny{H}}} = 0\, , \quad \mb^{(\e)}\ket{0}_{\mbox{\tiny{B}}} =
0\, , \quad (\om \e > 0)\, .\label{ETE63}
\end{equation}
According to Eq.~(\ref{ETE60}), these vacuum states are formally linked by
\begin{equation}\label{ETE64}
\ket{0}_{\mbox{\tiny{H}}} = e^{-iG}\, \ket{0}_{\mbox{\tiny{B}}}\, .
\end{equation}
This last relation needs to be treated with some caution, because strictly $\ket{0}_{\mbox{\tiny{H}}}$ and $\ket{0}_{\mbox{\tiny{B}}}$ are unitarily inequivalent \cite{ETE1}.

A more explicit form of the Eq.~(\ref{ETE64}) can be obtained by factorizing the operator $\expo{-iG}$ into its creation and annihilation operators, which is done by a generalization of the Baker-Campbell-Hausdorff (BCH) identity (see Appendix B):
\begin{equation}\label{ETE68}
e^{-iG_{\Om}}\, \ket{0}_{\mbox{\tiny{B}}} = Z_{\Om}^{-\mitad}\sum_{n_{\Om} =
0}^{\infty}e^{-\mitad\, \beta\, n_{\Om}\, \om}\, \ket{n_{\Om}^{(+)},
n_{\Om}^{(-)}}_{\mbox{\tiny{B}}}\, , 
\end{equation}
where $\ket{n_{\Om}^{(+)},n_{\Om}^{(-)}}_{\mbox{\tiny{B}}}$ is the Boulware state with an equal number $n_{\Om}$ of correlated Boulware modes in the state of the field $\Om$ ( i.e., it specified for the quantum numbers $\Om = \om, k, l, m$ with $\om > 0$) in the $L$ and $R$ sectors. Moreover: $\beta$ was introduced in terms of Hawking temperature $T_{\chiqui{H}}$, such that
\begin{equation}\label{ETE69}
\beta^{-1} = T_{\mbox{\tiny{H}}} = \frac{\kappa_{\chiqui{o}}}{2\pi}\, ,
\end{equation}
Eq.~(\ref{ETE31}) was rewritten as
\begin{equation}\label{ETE70}
\tanh \chi = e^{-\frac{\pi |\om|}{\kappa_{\chiqui{0}}}} = e^{-\mitad \beta \om}\, , \quad \om
> 0\, ,
\end{equation}
and we have used
\begin{equation}\label{ETE71}
\cosh^{2}\chi = \frac{1}{1 - e^{-\beta \om}} = \sum_{n=0}^{\infty}e^{-n\, \beta\,
\om} \equiv Z_{\Om}\, . 
\end{equation}
According to equations \eqref{ETE61} and \eqref{ETE68}, the expression \eqref{ETE64} can be reduced to 
\begin{align}
\ket{0}_{\mbox{\tiny{H}}} &= \prod_{\substack{\Om \\ \om >
0}}e^{-iG_{\Om}}\, \ket{0}_{\mbox{\tiny{B}}}\, ,\label{ETE72}\\
\ket{0}_{\mbox{\tiny{H}}} &= Z^{-\mitad}\sum_{\bn}e^{-\mitad\, \beta\,
E_{\bn}}\, \ket{\bn^{(+)}, \bn^{(-)}}_{\mbox{\tiny{B}}}\, ,\label{ETE73}
\intertext{where}
\bn &\equiv \{n_{\Om}\, , \quad \text{for all $\Om$ with $\om > 0$}\}, \notag \\
E_{\bn} &= \sum_{\substack{\Om \\ \om > 0}}n_{\Om}\, \om\, ,\quad
Z=\sum_{\bn}e^{-\beta E_{\bn}} = \prod_{\substack{\Om \\ \om > 0}}Z_{\Om}\,
. \label{ETE74}
\end{align}

Then, the ground state on the full Kruskal manifold is a thermal entangled state.
Vacuum state $\ket{0}_{\mbox{\tiny{B}}}$ has been written as a global state, but it does not really exist as a global state defined over the  $L + R$ region, because (KB)-modes are not positive frequency in any globally defined regular time parameter. The state $\ket{0}_{\mbox{\tiny{B}}}$ is empty of  (KB)-modes $\metedos^{(\e)}$ positive frequency in (future directed) Killing time $t$ in each of the sectors  $L$, $R$.

Consider the parochial  Boulware state $\ket{0}_{\mbox{\tiny{BR}}}$ defined over the complete Kruskal space:\\
from expression \eqref{ETE73}, it is clear that  
\begin{equation}\label{ETE76}
\ket{0}_{\mbox{\tiny{H}}} = Z^{-\mitad}\sum_{\bn}e^{-\mitad\, \beta\, E_{\bn}}\,
\ket{\bn}_{\mbox{\tiny{BL}}}\, \ket{\bn}_{\mbox{\tiny{BR}}}\,,
\end{equation}
where, $\ket{\bn}_{\mbox{\tiny{BL}}}$ and $\ket{\bn}_{\mbox{\tiny{BR}}}$ are excitations of the parochial Boulware states $\ket{0}_{\mbox{\tiny{BL}}}$ and $\ket{0}_{\mbox{\tiny{BR}}},$ respectively.

On the other hand, From Eq.~(\ref{ETE76}) we can think of $\ket{0}_{\mbox{\tiny{BR}}}$ as $\ket{0}\chiqui{H}$ depopulated of all Boulware modes $\ket{\bn}\chiqui{BR}$ in $R$ sector.
The parochial Boulware state $\ket{0}_{\mbox{\tiny{BR}}}$ is  empty of positive frequency Killing modes  $\metedos^{(+)}$ in the sector $R$:
\begin{equation}\label{ETE79}
\mb^{(+)}\, \ket{0}\chiqui{BR} = 0\, , \quad \om > 0\, . 
\end{equation} 

\subsection{Thermal energy of black shells}

Consider the expectation value for the component $T_{\chiqui{00}}$ of the stress-energy tensor with respect to Boulware and Hartle-Hawking states for the scalar field defined over metric \eqref{ETE1}.
In general, for the scalar field $\Phi$, the expectation value of the stress-energy tensor  $T_{\alpha \beta}(x)$ is given by
\begin{align}
\langle\, T_{\alpha \beta}(x, x')\, \rangle &= \mathcal{D}_{\alpha \beta'}\,
W(x, x')  ,\label{ETE104} \\
\langle\, T_{\alpha \beta}(x)\, \rangle &= \lim_{x \to x'} \langle\, T_{\alpha \beta}(x, x')\, \rangle, \notag
\end{align}
where
\begin{equation}
T_{\alpha \beta} = \Phi,_{\alpha}\, \Phi,_{\beta} - \mitad\, g_{\alpha
\beta}\left(\Phi'^{\gamma}\, \Phi,_{\gamma} + m^{2}\Phi^{2}\right)\,
,\label{ETE105}
\end{equation}
for the minimal coupling;
\begin{equation}
\mathcal{D}_{\alpha \beta'} = \partial(_\alpha\, \partial_{\beta'}) -
\mitad\, g_{\alpha \beta}\left(\partial^{\gamma}\partial_{\gamma'} +
m^{2}\right)\, ,\label{ETE106}
\end{equation}
and $W(x, x')$ is the Wightman function
\begin{equation}
W(x, x') = \espe{0}{\Phi
(x)\, \Phi(x')}{0}\, .\label{ETE107}
\end{equation}
 \eqref{ETE104} \,\,\,\,\,\,and \,\,\,\,\,\, \eqref{ETE107}\,\,\,\,\,\, become infinite when \,\,\,\,\,\,$x'\to x$, \,\,\, because mode sum
 \\ $\int_{0}^{\infty} d\om\,\varphi_{\Om}^{*}(x')\,\metedos(x)|_{x'=x}$ diverges as $\om \to \infty$. A regularization is required, and \,\,\,involves\,\,\, subtracting \,\,\,the \,\,\,(state-independent) Hadamard \,\,\,function \\$H(x, x')$ from $W(x, x')$ \cite{ELE}. But this will not be needed here, because we are only interested in differences between Wightman functions associated to Boulware and Hartle-Hawking states, and   $H(x, x')$ cancels out of Eq.~(\ref{ETE118}) and the corresponding $\langle\, T_{\alpha \beta} \rangle$.

Regarding Boulware state $\ket{0}\chiqui{B}$, the Wightman function is given by 
\begin{align}
W\chiqui{B}(x, x') &= \,{\chiqui{B}}\espe{0}{\Phi(x)\,
\Phi(x')}{0}{\chiqui{B}} \notag \\
&= \, \chiqui{B}\bra{0}{\sum_{\e, \Om, \e', \Om'}\mb^{(\e)}\,
\metedos^{(\e)}(x)\, b_{\Om'}^{(\e')\dag}\,
\varphi_{\Om'}^{(\e')*}(x')}\ket{0}{\chiqui{B}}, \label{ETE109}
\end{align}
where  expansion  \eqref{ETE59-1} was used.
Thus, by resorting to the result
\begin{align}
&\chiqui{B}\espe{0}{\mb^{(\e)}\, b_{\Om'}^{(\e')\dag}}{0}{\chiqui{B}} =
\Theta(\e\, \om)\, \delta_{\e \e'}\, \delta_{\Om \Om'}\, ,\label{ETE110}\\
\intertext{we find}
&\qquad W\chiqui{B}(x,x') = \sum_{\e, \Om} \Theta(\e\, \om)\,
\metedos^{(\e)}(x)\, \metedos^{(\e)*}(x')\, .\label{ETE111}
\end{align}

Since $\metedos^{(\e)}(x)$ is proportional to $\Theta_{\e}(x)$, given by Eq.~(\ref{ETE17}), $W\chiqui{B}(x, x') = 0$ if $x, x'$ are in opposite sectors
$L$, $R$.\\

Similarly it is obtained for $\ket{0}\chiqui{H}$, 
\begin{align}
W\chiqui{H}(x,x') &= \,\chiqui{H}\espe{0}{\Phi(x)\,
\Phi(x')}{0}{\chiqui{H}} \notag \\
&= \sum_{\e, \Om}\Theta(\e\, \om)\, \mkhh^{(\e)}(x)\, \mkhh^{(\e)*}(x')\,
.\label{ETE112}
\end{align}

By using the Bogolubov transformation \eqref{ETE30}, $W\chiqui{H}(x, x')$ can be expanded as
\begin{equation}\label{ETE113}
W\chiqui{H}(x,x') =\sum_{\e, \Om}\Theta(\e\, \om)\left[\cosh^{2}\chi\,
\metedos^{(\e)}(x)\, \metedos^{(\e)*}(x') 
   + \sinh^{2}\chi\,
\metedos^{(-\e)}(x)\, \metedos^{(-\e)*}(x')\right],
\end{equation}
for $x$, $x'$ in same sector, since $\metedos^{(\e)}(x)\,
\metedos^{(-\e)}(x') \propto \Theta_{\e}(x)\, \Theta_{-\e}(x') = 0$, when
$x$, $x'$ belong to same sector.

From the expressions \eqref{ETE113} and \eqref{ETE111}
\begin{align}
&W\chiqui{H}(x, x') - W\chiqui{B}(x, x') = (W\chiqui{H} - W\chiqui{B})(x,
x')  \notag \\
& \,\,\,= \sum_{\Om}\sinh^{2}\chi\left[\metedos^{(+)}(x)\,
\metedos^{(+)*}(x') + \metedos^{(-)}(x)\, \metedos^{(-)*}(x')\right],
\label{ETE116}
\end{align}
for $x$, $x'$ in same sector and where one used the identity
\begin{equation}
 \Theta(\om) + \Theta(-\om) = 1,
 \notag 
 \end{equation}
 \begin{align}
&(W\chiqui{H} - W\chiqui{B})(x, x')  \notag \\
&= \,\, \sum_{\Om}\frac{1}{e^{\beta|\om|} - 1}\, \left[\metedos^{(+)}(x)\,
\metedos^{(+)*}(x') + \metedos^{(-)}(x)\, \metedos^{(-)*}(x')\right],
\label{ETE117}
\end{align}
where it was introduced
\begin{equation}\notag
 \sinh^{2}\chi = \frac{\tanh^{2}\chi}{1 - \tanh^{2}\chi}\, ,
 \end{equation}
for $\tanh \chi$ defined by Eq.~(\ref{ETE31}).

Due to properties of $\metedos^{(\e)}(x)$ and $\metedos^{(-\e)}(x)$ it is necessary to restrict the calculations to sector $R$ or sector $L$. Then, if we suppose $x$, $x'$ belong to same sector, say $R$, finally  Eq.~(\ref{ETE117}) becomes
\begin{align}
&(W\chiqui{H} - W\chiqui{B})(x, x') \notag \\
&\qquad= \sum_{\Om}\frac{1}{e^{\beta |\om|} - 1}\metedos^{(+)}(x)\,
\metedos^{(+)*}(x')\, ,\label{ETE118}
\end{align}
because in sector $R$, $\Theta_{-} = 0$ and $\Theta_{+} = 1$,\\
then
\begin{align}
\metedos^{(-)}(x)\, \metedos^{(-)*}(x') \propto
\Theta_{-}(x)\, \Theta_{-}(x') =& 0\, ,\notag \\
 \metedos^{(+)}(x)\, \metedos^{(+)*}(x') \propto \Theta_{+}(x)\,
\Theta_{+}(x') =& 1\, . \notag 
\end{align}

To calculate $T_{\chiqui{00}}(x, x')$, first we evaluate 
\begin{equation}\label{ETE119}
\left.\partial_{\chiqui{0}}\, \partial_{\chiqui{0'}}(W\chiqui{H} - W\chiqui{B})(x, x')\right|_{x'
= x} \equiv \left. \partial_{\chiqui{0}}\, \partial_{\chiqui{0'}}\, W\chiqui{H-B}(x,
x')\right|_{x' = x},
\end{equation}
\begin{eqnarray}
\left. \partial_{\chiqui{0}}\, \partial_{\chiqui{0'}}\, W\chiqui{H-B}(x, x')\right|_{x' = x}
=&& \sum_{\Om}\frac{1}{e^{\beta |\om|} - 1} \notag \\ 
&& \times \partial_{\chiqui{0}}\, \partial_{\chiqui{0'}}\,
\left. \metedos^{(+)}(x)\, \metedos^{(+)*}(x')\right|_{x' = x} \notag \\
=&& \sum_{\Om}\frac{\mitad\, |\om|}{e^{\beta |\om|} - 1}\,
\left|\metedos(\bx)\right|^{2}\, ,\label{ETE120}
\end{eqnarray}
where Eq.~(\ref{ETE15}) was considered.\\
The expression \eqref{ETE120} can be reduced to 
\begin{eqnarray}
\left.\partial_{\chiqui{0}}\, \partial_{\chiqui{0'}}\, W\chiqui{H-B}(x, x')\right|_{x' = x} =&&
\frac{1}{4\pi}\int_{0}^{\infty}d\om\, \frac{\om}{e^{\beta \om} - 1} \notag \\
&& \times \sum_{l}(2l
+ 1)\left|\varphi_{\om\, l}(r)\right|^{2}, \nonumber \\
\label{ETE121}
\end{eqnarray}
by using
\begin{equation}
\metedos(\bx) = \varphi_{\om\, l}(r)\, Y_{l\, m}(\theta, \varphi)\, ,
\label{ETE122}
\end{equation}
\begin{equation}
\sum_{m = -l}^{l} \left|Y_{l\, m}(\theta, \varphi)\right|^2 = \frac{(2l +
1)}{4\pi}\, ,  \label{ETE123}
\end{equation}
\begin{equation}
\text{and} \quad \varphi_{\om l} = \varphi_{|\om|\, l,}\, \quad \text{then}\quad
\int_{-\infty}^{\infty}d\om\, () = 2\int_{0}^{\infty}d\om\, ().\label{ETE124}
\end{equation}
The result \eqref{ETE121} does not diverge since due to the difference between Boulware and Hartle-Hawking states introduced, the Bogolubov transformation \eqref{ETE30} supplies a convergence factor $(e^{\beta|\om|} - 1)^{-1}$.

To complete the calculation of  Eq.~(\ref{ETE121}) we need to known explicitly the function $\varphi_{\om l}$. Thus, in terms of the metric \eqref{ETE1A} and introducing the new function $\psi(r)$ in the mode solutions \eqref{ETE122} such that
\begin{equation}\label{ETE126}
\varphi(r) = \frac{1}{\sqrt{r^{2}\, f}}\, \psi(r)\, ,
\end{equation}
the equation Eq.~(\ref{ETE2*}) leads to the differential equation
\begin{equation}\label{ETE128}
\left[\dtotal{}{r} + k^{2}(r; \om, l)\right]\psi_{\om\, l}(r) = 0\,,
\end{equation}
where
\begin{equation}\label{ETE129}
k^{2} \equiv \frac{1}{f}\left\{\frac{\om^{2}}{f} - \frac{l(l + 1)}{r^{2}} -
m^{2} - \frac{(r^{2}\, f)''}{2r^{2}}\right\}\, ,
\end{equation}
and  the event horizon is characterized by $r = r_{\chiqui{0}}$, $f(r_{\chiqui{0}}) =
0$.

In order to solve the equation Eq.~(\ref{ETE128}) we resort to WKB approximation, then 
\begin{equation}\label{ETE130}
\psi_{\om\, l\, \eta}(r) \approx
\sqrt{\frac{1}{2\pi}\left|\frac{\om}{k(r)}\right|}\,
e^{i\int_{r_{\chiqui{0}}}^{r}k(r')\, dr'}\, ,
\end{equation}
where the new index $\eta=\sign k(r)$ corresponds to outgoing and ingoing waves: $\e\, \eta = \pm 1,$ which means that we need include $\eta$ in collective index  $\Om \equiv \om\, l\, m\, \eta$.

Before continuing the calculation of the expression \eqref{ETE119}, it is important to consider the normalization of WKB mode \eqref{ETE130}. So, spatial modes $\metedos(\bx)$ are required to satisfy
\begin{equation}\label{ETE131}
 \int d^{3}x\, \gamma(\bx)\, \varphi_{\Om'}^{*}(\bx)\,
\metedos(\bx) = \delta_{\Om \Om'} + \delta_{\bar{\Om} \Om'},
\end{equation}
where
\begin{equation}
 \Om = \om lm \eta, \,\, \bar{\Om} = -\om,lm\eta, \nonumber 
 \end{equation}
 \begin{equation}\label{ETE132}
 \gamma(\bx) =
 \sqrt{-g} (-g^{\chiqui{00}}) = f^{-1} r^{2}\sin \theta.
\end{equation}
In these terms, now we have
\begin{align}
 \metedos(\bx) =& \vphi_{\om l\eta}(r)\,
Y_{lm}(\theta, \vphi)\, , \nonumber \\
 \vphi_{\om l\eta}(r) =&
\frac{1}{\sqrt{r^{2}f}}\, \psi_{\om l\eta}(r)\, ,\label{ETE133}
\end{align}
\begin{equation}\label{ETE134}
 \iint d\theta\, d\vphi\, \sin \theta\, Y_{l'm'}^{*}(\theta,
\vphi)\, Y_{lm}(\theta, \vphi) = \delta_{ll'}\, \delta_{mm'}\, .
\end{equation}
Hence Eq.~(\ref{ETE131}) reduces to
\begin{equation}\label{ETE135}
\int dr f^{-2} \psi_{\om' l'\eta'}^{*}(r) \psi_{\om l\eta}(r) =
\left\{\delta(\om - \om') + \delta(\om + \om')
\right\}\delta_{\eta \eta'}.
\end{equation}

Returning to the calculation of \eqref{ETE119}, the expressions \eqref{ETE130} and \eqref{ETE133} are substituted into Eq.~(\ref{ETE121}) to obtain 
\begin{align}
&\partial_{\chiqui{0}}\, \partial_{\chiqui{0'}}\, W\chiqui{H-B} =
\frac{1}{4\pi^{2}r^{2}}\int_{0}^{\infty}\frac{\om^{2}\, d\om}{e^{\beta \om} -
1}\, \int_{0}^{l\chiqui{\text{max}}} \frac{(2l + 1)dl}{f|k|}\, ,\label{ETE137}
\intertext{where we approximate, for large $l$,} \notag
&\qquad \qquad \sum_{l}() \approx \int ()\,dl\, ,\quad \text{also}\quad
\sum_{\eta = \pm}() = 2\,().
\intertext{Moreover, $l\chiqui{\text{max}}(\om, r)$ was defined according to} 
&\qquad \qquad \qquad \qquad k^{2}(r; \om, l\chiqui{\text{max}}) = 0\, ,
\intertext{that is ,} 
&\qquad r^{-2}\, l\chiqui{\text{max}}(l\chiqui{\text{max}} + 1) =
\frac{\om^{2}}{f} - m^{2} - \frac{(r^{2}f)''}{2r^{2}} \equiv p^{2}\,
.\label{ETE138}\\
&\qquad \int_{0}^{l\chiqui{\text{max}}}\frac{(2l + 1)\, dl}{f|k|} =
2f^{-1}\sqrt{f}\, r\, \sqrt{l\chiqui{\text{max}}(l\chiqui{\text{max}} + 1)}\,
.\label{ETE139}
\end{align}

Finally, substituting Eq.~(\ref{ETE139}) into Eq.~(\ref{ETE137}) and multiplying by $(-g^{\chiqui{00}})$, we obtained 
\begin{align}
&-\partial^{\chiqui{0}}\, \partial_{\chiqui{0'}}\, W\chiqui{H-B} =
\frac{1}{2\pi^{2}}\int_{0}^{\infty}\frac{p\, \om^{2}\, d\om}{e^{\beta \om} -
1}\, \frac{\sqrt{f}}{f^{2}}\, . \label{ETE140} \\
&-\partial^{\chiqui{0}}\, \partial_{\chiqui{0'}}\, W\chiqui{H-B} =
\int_{0}^{\infty}\frac{E}{e^{\frac{E}{T}} - 1}\, \frac{4\pi p^{2}\,
dp}{(2\pi)^{3}}\, , \label{ETE142}
\intertext{where the local proper energy per mode $E$ has been defined by $E =
\frac{\om}{\sqrt{f}}$. Then, $\beta \om = \frac{E}{T(r)}$,}
&\qquad \qquad \quad T(r)\, \sqrt{-g_{00}} = T\chiqui{H} = \beta^{-1}\,
,\label{ETE141}\\
\intertext{which is Tolman's law, and}
&\qquad \qquad \quad p\, dp = \frac{\om\, d\om}{f} = E\, dE\, . \notag
\end{align}
Regarding the extra term in $-T_{\chiqui{0}}^{\chiqui{0}}$ arising from $\partial^{\gamma}\, \partial_{\gamma'} + m^{2}$ in Eq.~(\ref{ETE106}), can be shown to be zero to same approximation.
Then,
\begin{eqnarray}
\lim_{x \to x'} \langle\, T_{\chiqui{0}}^{\chiqui{0}}(x, x')\, \rangle\chiqui{H-B} &&=
\partial^{\chiqui{0}}\, \partial_{\chiqui{0'}}\, W\chiqui{H-B} \nonumber \\
&&= -\int_{0}^{\infty}\frac{E}{e^{\frac{E}{T}} - 1}\, \frac{4\pi\, p^{2}\,
dp}{(2\pi)^{3}}\, . \notag \\
\label{ETE153}
\end{eqnarray}
This is the thermal expression for energy density of a hot scalar field corresponding to eq.~(\ref{EU16-15}).

\subsection{Entanglement entropy of black shells}

To calculate the entanglement entropy $S$ consider the density matrix $\rho \chiqui{H}$ obtained from  $Eq.~(\ref{ETE76})$ and the reduced density matrix $\rho^{(+)}$,  given by
\begin{align}
\rho \chiqui{H} &= \ket{0}{\chiqui{H}}\, \chiqui{H}\bra{0}\, ,\label{ETE101}\, \\
\rho^{(+)} &= \text{tr}_{B^{(-)}}\, \rho \chiqui{H}\, ,\label{ETE100}
\end{align}
where the trace is taken over the degrees of freedom corresponding to sector $L$.\\
Thus, we may calculate entropy as
\begin{equation}
S = - \text{tr} \,( \rho^{(+)} \, \ln\, \rho^{(+)} ), \label{ETE100-1}
\end{equation}
with the reduced density matrix
\begin{equation}\label{ETE102}
\rho^{(+)} = Z^{-1}\sum_{\bn}e^{-\beta E_{\bn}}\, \ket{\bn^{(+)}}{\chiqui{B}}\,
\chiqui{B}\bra{\bn^{(+)}}\, ,
\end{equation}
where $Z$ is expressed by \eqref{ETE74}, and 
\begin{equation}\label{ETE103}
Z_{\Om} = \sum_{n = 0}^{\infty}e^{-n\beta \om} = \frac{1}{1 - e^{-\beta \om}}\,.
\end{equation}

The thermal system described above and associated to entanglement entropy actually allow us to think that entropy arises physically located near the horizon. Thus, we can calculate it from the expressions \eqref{ETE74} and \eqref{ETE103} in terms of the partition function $Z$:
\begin{equation}
S = -\beta \parcial{}{\beta}\, \ln Z + \ln Z\, ,\label{ETE154}
\end{equation}
which corresponds to
\begin{equation}
S = \beta\, U + \ln Z\, ,\label{ETE155}
\end{equation}
with
\begin{equation}
U = Z^{-1}\sum_{\bn}E_{\bn}e^{-\beta E_{\bn}} = -\parcial{}{\beta}\ln Z\,
.\label{ETE156}
\end{equation}

Since, for any function $f(\om)$ which goes to zero as  $\om \to\infty$, one can write,
\begin{align}
\sum_{\Om} f(\om) &\equiv \int_{0}^{\infty} d\om\, N(\om)\, f(\om)\,
,\label{ETE157}
\intertext{then}
\ln Z = \sum_{\Om}\ln Z_{\Om} &= \sum_{\Om}f(\om) = \int_{0}^{\infty}d\om\,
N(\om)\, f(\om)\, ,\label{ETE158}\\
\intertext{where}
f(\om) &= \ln \left(\frac{1}{1 - e^{-\beta \om}}\right)\, ,\label{ETE159}
\end{align}
and $N(\om)\, d\om$ is the number of modes $\metedos(x^{\alpha})$ which fall in the range $(\om, \om + d\om)$ for all admissible $k$, $l$, $m$.\\
In other words, in order to calculate $S$ associated to $\rho^{(+)}$ given by Eq.~(\ref{ETE100}), all what we need is to find $N(\om)$ finite, i.e., count modes:
\begin{equation}\label{ETE160}
\ln Z = \sum_{l, n, \eta = \pm}(2l + 1)\ln \left(\frac{1}{1 - e^{-\beta
\om_{ln}}}\right)\, . 
\end{equation}
Hence, to get $S$ finite we need to restrict the discrete sets $\{l = 0, 1, 2, \cdots\}$ and $\{n = 1, 2, \cdots\}$, which for the shell model is satisfied by the following restriction to the sets $\{l\}$ and $\{n\}$:
the discrete set $\{\eta = \pm 1;\, l = 0, 1, 2, \cdots;\, m = -l, \cdots,+l\}$ is restricted by the condition for the radial wave number $k$, defined by Eq.~(\ref{ETE129}), such that  
\begin{equation}\label{ETE161}
k^{2}(r; \om, l) \geq 0\, ,
\end{equation}
that is, $k$ is real; otherwise $\vphi(r)$ decays exponentially and is effectively zero. Hence, for a given $\om$, $N(\om)$ is finite or can be made finite by suitable boundary conditions.\\
In that sense, consider a finite-size confinement for the entanglement system, according to brick wall model, with Dirichlet boundary condition $\psi(r) = 0$ at inner boundary $r = r_{\chiqui{0}} + \e$.

WKB solution of the differential equation \eqref{ETE128} is the following expression

\begin{align}
&\psi_{ln}(r) = \frac{1}{\sqrt{k(r; \om_{ln}, l)}}\, \sin \int_{r_{0} +
\e}^{r_{ln}}k(r'; \om_{ln}, l)\, dr'\,, \label{ETE162}
\intertext{where for an admissible mode, both $k(r)$ and $\psi(r)$ must vanish simultaneously at the $n^{th}$ node $r = r_{n}$:}
&\qquad \qquad \qquad \int_{r_{0} + \e}^{r_{ln}}k(r'; \om_{ln}, l)\, dr' =
n\pi\, , \label{ETE163}\\
&\qquad \qquad \qquad k^{2}(r_{ln}, \om_{ln}, l) = 0\, , \label{ETE164}\\
&\qquad \qquad \qquad k^{2}(r'; \om_{ln}, l) \geq 0\, ,r'<r_{ln}, \label{ETE165}
\end{align}
which means that $k(r)$ is large near $r_{\chiqui{0}} + \e$ and decreases with increasing $r$. If $\frac{l(l + 1)}{r_{0}^{2}} > \om^{2}$, then $k^{2}(r)$ must eventually become zero; say, after $n$ oscillations of $\psi(r)$, $r = r_{n}$.\\
Since $\frac{\om_{ln}^{2}}{f(r)}$ in Eq.~(\ref{ETE129}) is very large near the inner wall, high-$l$ allowed modes are very numerous. So these modes are confined entirely to a thin layer near the inner wall, with the contribution
\begin{equation}\label{ETE166}
\sim \frac{l(l + 1)}{r_{\chiqui{0}}^{2}}\, .
\end{equation}
This dominant inner-mode contribution to $\ln Z$ and $S$ is proportional to  $r_{\chiqui{0}}^{2}$. On the other hand, outer modes (which satisfy boundary conditions at outer wall $r = R$) give a contribution proportional to $R^{3}$. This basically explains why the detailed calculations below gives
\begin{align}
S &= S\chiqui{wall} + S\chiqui{volume}\, ,\label{ETE168}
\intertext{where}
S\chiqui{wall} &\propto \, \text{horizon area}\, .\label{ETE167}
\end{align}

From Eq.~(\ref{ETE160}) we can write
\begin{equation}\label{ETE169}
\ln Z = \sum_{l, n}(2l + 1)\, \ln \left(\frac{1}{1 - e^{-\beta
\om_{ln}}}\right)\, ,
\end{equation}
where it was considered that Dirichlet boundary condition is imposed at inner boundary, o alternatively Neumann boundary condition, which does not affect counting.

Since $l,n$ are integers, the sum over $l$ and $n$ can be replaced to a good approximation by a double integral. Instead of $l,n$ as independent variables of integration, we switch to $l,\om$ as independent variables with $n=n(\om,l)$, and use the expressions \eqref{ETE163} and \eqref{ETE164} to write 
\begin{equation}\label{ETE170}\\
n(\om, l) = \frac{1}{\pi}\int_{r_{0} + \e}^{r(\om, l)}k(r'; \om, l)\, dr'\,
; k^{2}(r(\om, l); \om, l) = 0.
\end{equation}
\begin{eqnarray}
\parcial{n(\om, l)}{\om} =&& \frac{1}{\pi}\int_{r_{0} + \e}^{r(\om,
l)}\parcial{k(r'; \om, l)}{\om}\, dr' \notag \\
&& + \frac{1}{\pi} \parcial{r(\om,
l)}{\om}\, k(r(\om, l); \om, l), \label{ETE171}
\end{eqnarray}
where the second term is zero by Eq.~(\ref{ETE170}).

Now, from the equations \eqref{ETE169} and \eqref{ETE171}
\begin{eqnarray}
\ln Z =&& \frac{1}{\pi} \iiint\limits_{(k^{2}(r'; \om, l) \geq 0)}dl\, d\om\,
dr'\, (2l + 1) \notag \\
&& \times \parcial{k(r'; \om, l)}{\om}\, \ln\left(\frac{1}{1 - e^{-\beta
\om}}\right), \label{ETE172}
\end{eqnarray}
where the limits of integration are the surfaces $k^{2}(r'; \om, l) =
0$, $r' = r_{\chiqui{0}} + \e$, $r' = R$, with $l \geq 0$ and $\om \geq 0$ restricted only by the condition $k^{2}(r'; \om, l) \geq 0$.\\
Integrating Eq.~(\ref{ETE172}) by parts, it is obtained
\begin{equation}\label{ETE173}
\ln Z = \frac{1}{\pi} \iiint dr'\, (2l + 1)\, dl\, d\om\, k(r'; \om, l)\,
\frac{\beta}{e^{\beta \om} - 1}\, .
\end{equation}
On the other hand,
\begin{align}
\int (2l + 1) dl\, k(r'; \om, l) &=
\frac{1}{r'\sqrt{f}}\int_{0}^{L\chiqui{max}(\om, r')} \mbox{\small $dL (L\chiqui{max} -
L)^{\mitad}$} \notag \\
&= \frac{1}{r'\sqrt{f}} \frac{2}{3} L\chiqui{max}^{\frac{3}{2}}\,
,\label{ETE174}
\end{align}
where 
\begin{equation*}
L \equiv l(
l + 1) \leq L\chiqui{max}(\om, r')\, ,
\end{equation*}
for $L\chiqui{max}$ given by the value that makes $k^{2} = 0;$ i.e., from Eq.~(\ref{ETE129})
\begin{equation}\label{ETE175}
L\chiqui{max}(\om, r') = \frac{r'^{2}}{f(r')}\left[\om^{2} - \left(m^{2} +
\frac{(r'^{2}\, f)''}{2r'^{2}}\right)f(r')\right]\, .
\end{equation}
Substituting Eq.~(\ref{ETE174}) into Eq.~(\ref{ETE173}) 
\begin{equation}\label{ETE176}
\ln Z = \frac{1}{\pi} \int_{0}^{\infty}d\om\, \int_{r_{0} + \e}^{R}dr'\,
\frac{\beta}{e^{\beta \om} - 1}\, \frac{1}{r'\sqrt{f}}\, \frac{2}{3}\,
L\chiqui{max}^{\frac{3}{2}}\, .
\end{equation}

From Eq.~(\ref{ETE176}) we can obtain usual thermodynamic expressions in terms of the Helmholtz free energy $F$ and the average energy $U$: 
\begin{equation}\label{ETE177}
F = -\frac{1}{\beta}\, \ln Z = -\int_{0}^{\infty} \frac{N(\om)\,
d\om}{e^{\beta \om} - 1},
\end{equation}
where
\begin{equation}\label{ETE178}
N(\om) = \frac{2}{3\pi}\int_{r_{0} + \e}^{R}\frac{r^{2}\, dr}{f^{2}(r)}\,
\left[\om^{2} - \left(m^{2} + \frac{(r'^{2}\
, f)''}{2r'^{2}}\right)f(r)\right]^{\frac{3}{2}}.
\end{equation}
\begin{eqnarray}
U =&& -\parcial{}{\beta}\, \ln Z = -\int_{0}^{\infty} d\om\,
N(\om) \nonumber \\
&& \times \left[\frac{1}{e^{\beta \om} - 1} - \beta\, \frac{e^{\beta
\om}}{(e^{\beta \om} - 1)^{2}}\, \om \right] \notag \\
 =&& -\int_{0}^{\infty}\frac{d\om}{e^{\beta \om} - 1}\left[N(\om) -
\parcial{}{\om}(\om\, N(\om))\right] \notag \\
 =&& \int_{0}^{\infty}\frac{d\om}{e^{\beta \om} - 1}\, \om\, N'(\om)\,
,\label{ETE179}
\end{eqnarray}
\begin{align}
\intertext{with $N'(\om)$ defined by}
&N'(\om) = \frac{2}{\pi}\, \om \int_{r_{0} + \e}^{R}\frac{r^{2}\,
dr}{f^{\frac{3}{2}}}\, \sqrt{\frac{\om^{2}}{f} - \left(m^{2} + \frac{(r'^{2}\,
f)''}{2r'^{2}}\right)}\, .\label{ETE180}\\
\intertext{Then, entropy $S$ gives }
&S = \beta(U - F) = \beta \int_{0}^{\infty} \frac{\om\, N'(\om) +
N(\om)}{e^{\beta \om} - 1}\, d\om\, .\label{ETE181}
\end{align}

To convert the expressions above to a statistical-thermodynamic form, we can change variables of integration from $r$ and $\om$ to $r$ and $p$, where
\begin{equation}\label{ETE182}
p^{2} = \frac{\om^{2}}{f(r)} - m
^{2} - \frac{(r^{2}\, f)''}{2r^{2}}\, .
\end{equation}
Then,
\begin{equation}
p\, dp\, dr = \frac{\om\, d\om}{f}\, dr\, .\label{ETE183}
\end{equation}
In these terms, the expressions \eqref{ETE179} and \eqref{ETE180} become
\begin{equation}\label{ETE184}
U = \int_{r_{0} + \e}^{R}4\pi\, r^{2}\, dr\, \int_{0}^{\infty}\frac{E(p,
r)}{e^{\beta \om} - 1}\, \frac{4\pi\, p^{2}\, dp}{(2\pi\, \hbar)^{3}}\,,
\end{equation}
where we have explicitly restored $\hbar = 1$ and introduced the locally measured energy of mode with frequency $\om$, $E = \frac{\om}{\sqrt{f}}$; with $\om(p, r)$ given by Eq.~(\ref{ETE182}), and the effective local momentum $p$.\\
Then,
\begin{equation}
\beta \om = \frac{E}{T(r)}\, , \quad T(r) = \frac{\beta^{-1}}{\sqrt{f}}\,
. \label{ETE185}
\end{equation}
Thus,
\begin{equation}
U = \int_{r_{0} + \e}^{R}4\pi\, r^{2}\, \rho(r)\, dr\, ,\label{ETE186}
\end{equation}
with
\begin{equation}
\rho(r) = \int_{0}^{\infty}\frac{E}{e^{\frac{E}{T}} - 1}\, \frac{4\pi\,
p^{2}\, dp}{h^{3}}\, ,\label{ETE187}
\end{equation}
\begin{equation}
E^{2} = p^{2} + m^{2} + \frac{(r^{2}\, f)''}{2r^{
2}} \approx p^{2} + m^{2}\, .\label{ETE188} 
\end{equation}
Finally, from Eq.~(\ref{ETE181})
\begin{align}
S &= \beta \int_{0}^{\infty} \parcial{}{\om} (\om\, N(\om))\,
\frac{1}{e^{\beta \om} - 1}\, d\om \notag \\
&= \beta^{2}\int_{0}^{\infty}\om\, N(\om)\, \frac{e^{\beta \om}}{(e^{\beta
\om} - 1)^{2}}\, d\om \, .\label{ETE189}
\end{align}

Now, resorting to the equations \eqref{ETE178}, \eqref{ETE182} and \eqref{ETE183}
\begin{align}
S &= \frac{2}{3\pi}\, \beta^{2}\iint \frac{r^{2}\, dr}{f^{2}}(p^{2}\,
f)^{\frac{3}{2}}\, \frac{e^{\beta \om}}{(e^{\beta \om} - 1)^{2}}\, p\, dp\,
f \notag \\
&= \frac{1}{3}\, \beta^{2}\int 4\pi\, r^{2}\, \frac{dr}{\sqrt{f}}\cdot f\int
\frac{p^{2}\, e^{\beta \om}}{(e^{\beta \om} - 1)^{2}}\, \frac{4\pi\, p^{2}\,
dp}{(2\pi \hbar)^{3}}\, .\label{ETE190}
\end{align}
Thus,
\begin{equation}
S = \int_{r_{0} + \e}^{R}4\pi\, r^{2}\, \frac{dr}{\sqrt{f}}\, s(r)\,
,\label{ETE191}
\end{equation}
where
\begin{equation}
s(r) = \frac{1}{3T^{2}}\int_{0}^{\infty} \frac{p^{2}\,
e^{\frac{E}{T}}}{(e^{\frac{E}{T}} - 1)^{2}}\, \frac{4\pi\, p^{2}\, dp}{h^{3}}\,
.\label{ETE192}
\end{equation}

The analysis above corresponds to brick wall model \cite{BW1,BW2}. According to this model, the integrals \eqref{ETE186} and \eqref{ETE191} are dominated by two contributions, for large $r=R$ and for small $r_{\chiqui{0}} + \e$. The former corresponds to a volume term, proportional to $\frac{4}{3}\pi r^{3}$, which represents the entropy and energy of a homogeneous quantum gas in a flat space at a uniform temperature $\frac{k_{0}}{2\pi}$. The latter is the contribution of gas near the inner wall $r=R_{\chiqui{0}}$. Then, for this last contribution is required to introduce the ultrarelativistic approximations
\begin{equation}\label{BW8}
s=\frac{4N}{\pi^{2}} T^{3} \text{,}\,\, \rho = \frac{3N}{\pi^{2}}T^{4}.
\end{equation}

Substituting Eq.~(\ref{BW8}) into Eq.~(\ref{ETE191}), the wall contribution to the total entropy is obtained:
\begin{equation}\label{BW10}
S_{wall} = \frac{N}{90\pi \alpha^{2}} \frac{1}{4}A,
\end{equation}
where $N$ accounts for helicities and the number of particle species,  $A$ is the wall area and $\alpha$ is the proper altitude of the inner wall above the horizon of the exterior geometry.

Now, depending on $\alpha$, we can obtain the Bekenstein-Hawking entropy from Eq.~(\ref{BW10}) 
\begin{equation}\label{BW12}
S_{wall} = S\chiqui{BH}, 
\end{equation}
where $\alpha$ has been adjusted by invoking quantum gravity effects.

\section{Conclusion}

We have presented an integrated and detailed explanation of how the entanglement interpretation, the thermofield description and the brick wall formulas (properly interpreted as referring to thermal excitations above the Boulware ground state) fit together to form a coherent, self-consistent explanation of what Bekenstein-Hawking entropy is, and where it is located. Nearly all the other brick wall papers are just formal calculations which, for instance, give no clue why the large entropies they derive near the horizon should sit on a region which is empty and nearly flat.

Really, in this paper we have introduced a model of black shell in terms of an effective Hartle-Hawking state, which is externally indistinguishable from a Schwarzschild black hole with respect to its thermodynamic properties. For this reason, the model considered above can be interpreted as an effective calculation of the entanglement entropy associated with a Schwarzschild black hole, i.e., $S\chiqui{BH}$ can be considered as entanglement entropy, which  is well defined near the horizon and presents a thermal nature according to thermofield dynamics of black holes.

On the other hand, by using thermodynamical arguments we suggest that the brick wall model might be considered as a model of black shell in order to get an operational approach  to black hole entropy~\cite{ref:P}. The topped-up Boulware state (TUB) defined there may be called a generalized Hartle-Hawking state more than an effective one. Indeed, it becomes the Hartle-Hawking state in the limit when the shell approaches its gravitational radius.

The most interesting interpretation with respect to the model of black shell developed above is just that maybe all existing derivations of $S\chiqui{BH}$ are ``superficial'', in the sense that they refer to an effective black shell entropy and don't probe the real black hole interior at all.  The viewpoint advocated in this paper suggests that it is possible to entertain the suspicion that all derivations of the Bekenstein-Hawking entropy formula, which differ so vastly in appearance, are just disguised variations of the same derivation. All derivations lead to the same formula because all calculate entropy of the same object: a black shell instead of a hole.\\

\textbf{\Large{Acknowledgments}}\\

We are indebted to Werner Israel, our collaborator of much of the work presented here, for many contributions, stimulating discussions and helpful comments on the manuscript.\\

\textbf{\Large{Appendix}}

\appendix

\section{Positive frequency modes}

\subsection{\label{app:subsec}Definition}

Let $\ln_{+}x$ be defined, for a real number $x$, by
\begin{align}
\label{eq:MFP1}
\ln_{+}x &\equiv \ln|x| + \frac{i\pi}{2}\epsilon(x)\, , \quad -\infty < x <
\infty \, ,\\
\intertext{where}
\e(x) &\equiv \sign (x)\, .\notag
\end{align}
Then,
\begin{equation}
e^{\pm i\alpha \ln_{+}x}
\end{equation}
are positive frequency functions in $x$ ( for both signs, $\pm
\alpha$, $\alpha$ real), i.e.  
\begin{equation}
e^{\pm i\alpha \ln_{+}x} = \int_{0}^{\infty} A_{\pm}(\omega)\, e^{-i\omega x}\,
d\omega\, .
\end{equation}
This assertion, whose proof is shown bellow,  it will be taken as definition of a positive frequency function.

\subsection{\label{app:subsec}Extension to an analytic function}

To prove the assertion above, extend the expression \eqref{eq:MFP1} to an analytic fuction $\lomas$ by defining:
\[ \lomas :
\begin{cases}
&\text{real on lower imaginary axis}\\
&\text{branch cut in upper half-plane}\, ,
\end{cases}
\]
that is, 
\begin{equation}
\lomas = \ln|z| + i(\text{arg}Z + \frac{\pi}{2})\, ,\,\,\,
-\frac{3\pi}{2} < \text{arg}z < \frac{3\pi}{2}\, .
\end{equation}
Then $\lomas$ is regular in lower half-plane.\\
$e^{\pm i\alpha \lomas}$  are regular and bounded in lower half-plane.\\
Hence
\begin{equation}
A_{\pm}(\omega) = \frac{1}{2\pi} \int_{-\infty}^{\infty} e^{\pm i\alpha
\lomas}\, e^{i\omega z}\, dz = 0 \quad \text{if} \quad \omega < 0\, .
\end{equation}
Finally it is defined
\begin{equation}
\ln_{\epsilon}x = \ln|x| + \frac{i\pi}{2}\epsilon(x)\, \epsilon\, , \quad
-\infty < x < \infty\, ; \quad \epsilon = \pm 1\, . 
\end{equation}
Then, 
\[ e^{\pm i\alpha \ln_{\epsilon}x} \, \text{are}
\begin{cases}
&\text{positive frequency in $x$ for $\epsilon = +1$}\\
&\text{negative frequency in  $x$ for $\epsilon = -1$}\, .
\end{cases}
\]

\subsection{\label{app:subsec}Application to maximally extended black hole geometry}

Let sectorial functions be defined by
\begin{equation}
\Theta_{
\epsilon}(x) \equiv \mitad \left\{\Theta(-\epsilon \, U) + \Theta(\epsilon \,
V)\right\}\, , 
\end{equation}
where $\Theta$ is the unit step function and $U,V$ the Kruskal times.\\ 
$U$ and $V$ can be defined by
\begin{equation}
T \pm Z =
\begin{Bmatrix}
V \\ U
\end{Bmatrix};
\end{equation}
\begin{equation}
\begin{Bmatrix}
\frac{dV}{\kappa_{\chiqui{0}}V} \\ \\ \frac{dU}{-\kappa_{\chiqui{0}}U}
\end{Bmatrix}
=
\begin{Bmatrix}
dv \\ du
\end{Bmatrix}
=
dt \pm \frac{dr}{f(r)}.
\end{equation}

Consider the definition
\begin{align}
2\kappa_{\chiqui{0}}\, t_{+} &\equiv \ln_{+}V - \ln_{+}U = \ln\left|\frac{V}{U}\right| +
\frac{i\pi}{2}(\epsilon(V) - \epsilon(U)) \notag\\
&= \ln\left|\frac{V}{U}\right| + i\pi(\Theta_{+} - \Theta_{-})\, ,
\end{align}
and the extension
\begin{align}
2\kappa_{\chiqui{0}}\, t_{\epsilon} &\equiv \ln_{\epsilon}V - \ln_{\epsilon}U = \ln\left|\frac{V}{U}\right| +
\frac{i\pi}{2}(\epsilon(V) - \epsilon(U))\epsilon \notag\\
&= \ln\left|\frac{V}{U}\right| + i\pi(\Theta_{\epsilon} - \Theta_{-\epsilon})\, ,
\end{align}
then
\[
e^{\pm i\alpha t_{\epsilon}}\, \text{are}
\begin{cases}
& \text{{\footnotesize positive frequency functions in $U$, $V$ for $\epsilon = +1$}}\\
& \text{{\footnotesize negative frequency functions in $U$, $V$ for $\epsilon = -1$}}\, ,
\end{cases}
\]

\subsection{\label{app:subsec}Useful relations}

\begin{align}
&(1)\quad \Theta_{\epsilon} + \Theta_{-\epsilon} = 1\, , \quad
\Theta_{\epsilon} - \Theta_{-\epsilon} = \mitad \, \epsilon \,
\left\{\epsilon(V) - \epsilon(U)\right\} \, .\\  
&(2) \quad e^{-i\omega t_{\epsilon \epsilon'}} = e^{-i\omega
t}\left(e^{\mitad \, \frac{\pi \omega}{\kappa_{\chiqui{0}}}\epsilon'}\Theta_{\epsilon} +
e^{-\mitad \, \frac{\pi \omega}{\kappa_{\chiqui{0}}}\epsilon'}\Theta_{-\epsilon}\right)\,
.\\
&(3) \quad \text{Let $\, \epsilon' = \epsilon(\omega)$, then} \notag\\
\label{eq:MFP12}
&\quad \quad e^{-i\omega t_{\epsilon(\omega)\, \epsilon}} = e^{-i\omega
t}\left(e^{\mitad \, \frac{\pi |\omega|}{\kappa_{\chiqui{0}}}}\Theta_{\epsilon} + e^{-\mitad
\, \frac{\pi |\omega|}{\kappa_{\chiqui{0}}}}\Theta_{-\epsilon}\right)\, .
\end{align}

Defining 
\begin{equation}
\chi = \chi(\omega)\quad \text{by} \quad  \tanh \chi = e^{-\frac{\pi
|\omega|}{\kappa_{\chiqui{0}}}}\,,  
\end{equation}
 Eq.~(\ref{eq:MFP12})  can be written as 
\begin{equation}\label{eq:MFP14}
e^{-i\omega t_{\epsilon(\omega) \epsilon}} = e^{-i\omega t}\left[
\left(\frac{\mbox{\small $\cosh \chi$}}{\mbox{\small $\sinh \chi$}}\right)^{1/2}\Theta_{\epsilon} +
\left(\frac{\mbox{\small $\sinh \chi$}}{\mbox{\small $\cosh \chi$}}\right)^{1/2}\Theta_{-\epsilon}\right]\, .
\end{equation}
 \begin{align}
[(\mbox{\small$\sinh \chi$})(\mbox{\small$\cosh \chi$})]^{\chiqui{1/2}}\,e^{-i\omega t_{\epsilon(\omega)\,
\epsilon}} =&e^{-i\omega t} \nonumber\\
 & \times \left[\mbox{\small$\cosh \chi$} \Theta_{\epsilon} + \mbox{\small$\sinh
\chi$} \Theta_{-\epsilon}\right], \notag \\
 \label{eq:MFP15}
\end{align}

\section{Baker-Campbell-Hausdorff identity}

If two operators $A$,$B$ and their commutator $C$ satisfy the commutation relations 
\begin{equation}\label{ETE65}
[A,B] = C\, ,\quad [C,A] = 2\,n^{2}\, A\, ,\quad [B,C] = 2\, n
^{2}\, B\, , 
\end{equation}
for some number $n$, real or complex, then for any parameter $\chi$,
\begin{equation}\label{ETE66}
e^{\chi(A+B)} = e^{\frac{1}{n}(\tanh \chi)A}\, e^{\frac{1}{2n}(\sinh 2n\chi)B}\,
e^{-\frac{1}{n^{2}}(\ln \cosh n\chi)C}\, .
\end{equation}

An outline of the derivation of the identity \eqref{ETE66} can be shown by introducing the operator \cite{ETE2}
\begin{equation}\label{BCH1}
F(\chi)=e^{x(\chi)A}\,e^{y(\chi)B}\,e^{z(\chi)C},\,
\end{equation}
where $x,y,z$ are undetermined functions. The idea of the proof consist in choosing these functions so that $F$ is reducible to $e^{\chi(A+B)}$.\\
Differentiating Eq.~(\ref{BCH1}) with respect to $\chi$,
\begin{equation}\label{BCH2}
\mbox{\small $F^{-1}F'(\chi)=x'e^{-zC}(e^{-yB}Ae^{yB})e^{zC} + y'e^{-zC}Be^{zC}+ z'C$},
\end{equation}
we obtain
\begin{equation}\label{BCH3}
\mbox{\footnotesize $F^{-1}F'(\chi)=x'(e^{-2mz}A+yC-y^{\chiqui{2}}me^{2mz}B)+y'e^{2mz}B+z'C$},
\end{equation}
where
\begin{eqnarray}\notag
e^{-yB}Ae^{yB}=&& A + y[A,B] + \frac{1}{2}y^{\chiqui{2}}[[A,B],B] + ... \nonumber \\
=&& A+yC-y^{\chiqui{2}}mB, \nonumber
\end{eqnarray} 
\begin{equation}\notag
e^{-zC}Ae^{zC}=e^{-2mz}A, \quad e^{-zC}Be^{zC}=e^{-2mz}B, 
\end{equation}
considering
\begin{equation}\notag
[C,A] = 2m A\, ,\quad [B,C] = 2m B. 
\end{equation}
Now, it is required Eq.~(\ref{BCH3})  to equal $A+B$. Equating coefficients of $A,B$ and $C$ yields three equations for $x,y,z:$
\begin{equation}\label{BCH4}
x'e^{-2mz}=1, \quad (y' -mx'y^{\chiqui{2}})e^{2mz}=1, \quad z'+x'y=0.
\end{equation}
If we define $f(\chi)=e^{-2mz}$, the first and third of these equations \eqref{BCH4} then give $x'=f^{-1}, y=f'/2m.$ Substituting into the second equation results in a second-order linear equation for $ f^{\chiqui{1/2}}.$ Finally, in order to find the generalized identity \eqref{ETE66}, consider the solution of this last linear equation, subject to the initial conditions $x \thickapprox y \thickapprox \chi, z \thickapprox 0$ when $\chi \to 0$ is $f(\chi)=\cosh^{\chiqui{2}} {n\chi}$.\\

Thus we can define
\begin{align*}
A_{\Om} &\equiv \mb^{(+)\dag}\, \mb^{(-)}\, , \quad B_{\Om} \equiv
-\mb^{(+)}\, \mb^{(-)\dag}\, , \quad \om > 0\, .
\intertext{Then,}
C_{\Om} &= \mb^{(+)\dag}\, \mb^{(+)} + \mb^{(-)\dag}\, \mb^{(-)}\, , 
\end{align*}
and the commutation relations \eqref{ETE65} are satisfied with $n = 1$, which means  that Eq.~(\ref{ETE66})  is applicable.\\

According to Eq.~(\ref{ETE61}), it is directly obtained 
\begin{align}
e^{-iG}\, \ket{0}_{\mbox{\tiny{B}}} &= e^{\chi(A_{\Om} + B_{\Om})}\,
\ket{0}_{\mbox{\tiny{B}}} \notag \\
&= \frac{1}{\cosh \chi}\, e^{(\tanh \chi)\mb^{(+)\dag}\, \mb^{(-)}}\,
\ket{0}_{\mbox{\tiny{B}}} \notag \\
&= \frac{1}{\cosh \chi}\sum_{n=0}^{\infty}(\tanh^{n} \chi)\ket{n_{\Om}^
{(+)},n_{\Om}^{(-)}}_{\mbox{\tiny{B}}}, \label{ETE67}
\end{align}
which corresponds to Eq.~(\ref{ETE68}).

\end{document}